%% file: maintext.tex
\pdfoutput=1
\documentclass[nature]{article}
\usepackage[margin=1in]{geometry}
\usepackage{graphicx}
\usepackage{epsfig}
\usepackage{fullpage}
\usepackage[super,sort&compress]{natbib}
\usepackage{epstopdf}
\usepackage{caption}
\newcommand{\spacing}[1]{\renewcommand{\baselinestretch}{#1}\large\normalsize}
\spacing{2}
\def\@maketitle{%
  \newpage\spacing{1}\setlength{\parskip}{12pt}%
    {\Large\bfseries\noindent\sloppy \textsf{\@title} \par}%
    {\noindent\sloppy \@author}%
}
\newenvironment{affiliations}{%
    \setcounter{enumi}{1}%
    \setlength{\parindent}{0in}%
    \slshape\sloppy%
    \begin{list}{\upshape$^{\arabic{enumi}}$}{%
        \usecounter{enumi}%
        \setlength{\leftmargin}{0in}%
        \setlength{\topsep}{0in}%
       \setlength{\labelsep}{0in}%
        \setlength{\labelwidth}{0in}%
        \setlength{\listparindent}{0in}%
        \setlength{\itemsep}{0ex}%
        \setlength{\parsep}{0in}%
        }
    }{\end{list}\par\vspace{12pt}}

\renewenvironment{abstract}{%
    \setlength{\parindent}{0in}%
    \setlength{\parskip}{0in}%
    \bfseries%
    }{\par\vspace{-6pt}}

\title{Circular polarization in the optical afterglow of GRB\,121024A}

\author{K. Wiersema$^1$, S. Covino$^2$, K. Toma$^{3,4,5}$, A. J. van der Horst$^6$, K. Varela$^7$,\\ 
M. Min$^6$, J. Greiner$^7$, R. L. C. Starling$^1$, N. R. Tanvir$^1$, \\ 
R. A. M. J. Wijers$^6$, S. Campana$^2$,  P. A. Curran$^8$, Y. Fan$^9$, \\
J. P. U. Fynbo$^{10}$, J. Gorosabel$^{11,12,13}$, A. Gomboc$^{14}$, D. G\"{o}tz$^{15}$, \\
J. Hjorth$^{10}$, Z. P. Jin$^9$, S. Kobayashi$^{16}$, C. Kouveliotou$^{17}$, \\
C. Mundell$^{16}$, P. T. O'Brien$^1$, E. Pian$^{18,19}$, A. Rowlinson$^6$, \\
D. M. Russell$^{20,21,22}$, R. Salvaterra$^{23}$, S. di Serego Alighieri$^{24}$, \\
G. Tagliaferri$^{2}$, S. D. Vergani$^2$,  J. Elliott$^7$, C. Fari\~{n}a$^{25}$, \\
O. E. Hartoog$^6$, R. Karjalainen$^{25}$, S. Klose$^{26}$, F. Knust$^7$, \\
A. J. Levan$^{27}$, P. Schady$^7$, V. Sudilovsky$^7$ \& R. Willingale$^1$ }

\begin{document}
\setcounter{page}{0}
\maketitle
\begin{affiliations}
\item Department of Physics \& Astronomy, University of Leicester, Leicester, LE1 7RH, United Kingdom
\item INAF/Brera Astronomical Observatory, via Bianchi 46, I-23807, Merate (LC), Italy
\item Department of Earth and Space Science, Osaka University, Toyonaka 560-0043, Japan 
\item Astronomical Institute, Tohoku University, Sendai 980-8578, Japan
\item Frontier Research Institute for Interdisciplinary Sciences, Tohoku University, Sendai 980-8578, Japan
\item Astronomical Institute `Anton Pannekoek', University of Amsterdam, PO Box 94248, 1090 SJ Amsterdam, the Netherlands
\item Max-Planck-Institut f\"{u}r extraterrestrische Physik, Giessenbachstrasse 1, D-85748 Garching,  Germany
\item International Centre for Radio Astronomy Research, Curtin University, GPO Box U1987,  Perth, WA 6845, Australia 
\item Key Laboratory of Dark Matter and Space Astronomy, Purple Mountain Observatory,  Chinese Academy of Science, Nanjing, 210008, China
\item Dark Cosmology Centre, Niels Bohr Institute, University of Copenhagen, Juliane Maries Vej 30, DK-2100 Copenhagen ¯, Denmark
\item Instituto de Astrof\'{\i}sica de Andaluc\'{\i}a (IAA-CSIC), Glorieta de la Astronom\'{\i}a s/n, E-18008, Granada, Spain
\item Unidad Asociada Grupo Ciencia Planetarias UPV/EHU-IAA/CSIC, Departamento de F\'{\i}sica Aplicada I, E.T.S. Ingenier\'{\i}a, Universidad del Pa\'{\i}s Vasco UPV/EHU, Alameda de Urquijo s/n, E-48013 Bilbao, Spain.
\item Ikerbasque, Basque Foundation for Science, Alameda de Urquijo 36-5, E-48008 Bilbao, Spain.
\item Faculty of Mathematics and Physics, University of Ljubljana, Jadranska 19, 1000 Ljubljana, Slovenia
\item AIM (UMR 7158 CEA/DSM-CNRS-Universit\'{e} Paris Diderot) Irfu/Service d'Astrophysique, Saclay, F-91191 Gif-sur-Yvette Cedex, France
\item Astrophysics Research Institute, Liverpool John Moores University, Liverpool Science Park,ÊIC2 Building, 146 Brownlow Hill, Liverpool L3 5RF, United Kingdom
\item Space Science Office, ZP12, NASA/Marshall Space Flight Center, Huntsville, AL 35812, USA
\item Scuola Normale Superiore, 7, I-56126, Pisa, Italy
\item INAF/IASF Bologna, via Gobetti 101, I-40129, Bologna, Italy
\item Instituto de Astrof\'{\i}sica de Canarias (IAC), E-38200 La Laguna, Tenerife, Spain
\item  Departamento de Astrof\'{\i}sica, Universidad de La Laguna, E-38206 La Laguna, Tenerife,  Spain
\item New York University Abu Dhabi, P.O. Box 129188, Abu Dhabi, United Arab Emirates
\item INAF/IASF Milano, via E. Bassini 15, 20133 Milano, Italy
\item INAF-Osservatorio Astrofisico di Arcetri, L.go E. Fermi 5, I-50125 Firenze, Italy
\item Isaac Newton Group of Telescopes, Apartado de Correos 321, E-38700 Santa Cruz de la Palma, Canary Islands, Spain
\item Th\"{u}ringer Landessternwarte Tautenburg, Sternwarte 5, 07778, Tautenburg, Germany
\item Department of Physics, University of Warwick, Coventry, CV4 7AL, United Kingdom 

\end{affiliations}
\begin{abstract}
Gamma-ray bursts (GRBs) are most probably powered by collimated relativistic outflows (jets) from accreting black holes at cosmological distances. Bright afterglows are produced when the outflow 
collides with the ambient medium. Afterglow polarization directly probes the magnetic properties of the jet, when measured minutes after the burst,  and the geometric 
properties of the jet and the ambient medium when measured hours to days after the burst$^{1,2,3,4,5}$. High values of optical polarization detected minutes after burst in 
GRB\,120308A indicate the presence of large-scale ordered magnetic fields originating from the central engine$^5$ (the power source of the GRB). 
Theoretical models predict low degrees of linear polarization and negligable circular polarization at late times$^{6,7,8}$, when the energy in the original ejecta is quickly transferred to the ambient medium and propagates farther into the medium as a blastwave. Here we report the detection of circularly polarized optical light in the afterglow of GRB 121024A, measured 0.15 days after the burst. We show that the circular polarization is intrinsic to the afterglow and unlikely to be produced by dust scattering or plasma propagation effects. A possible explanation is to invoke anisotropic (rather than the commonly assumed isotropic) electron pitch angle distributions, and we suggest that new models are required to produce the complex microphysics of realistic shocks in relativistic jets$^{9,10,11}$.
\end{abstract}

Magnetic fields play a crucial role in the physics of relativistic jets, for example in the jet formation, acceleration 
and collimation processes$^{12,13,1}$. On smaller spatial scales, there is a strong connection between particle 
acceleration and magnetic field generation in the collision-less relativistic shocks that create GRB afterglows$^{9,10,11}$. 
Our understanding of magnetic field properties in GRBs and their afterglows has improved rapidly through  
recent observational successes, such as time resolved linear polarimetry of prompt gamma-ray emission$^{14,15}$ (just seconds after burst) 
and the observed transition of reverse shock emission (high optical linear polarization, seen just minutes after burst)  to the early forward shock afterglow 
emission (lower levels of linear polarization)$^{5}$. These observations have given support to models predicting large scale ordered fields in the GRB ejecta$^{5}$.
 Late-time polarimetry (hours to days after burst) on the other hand offers the advantage that at these times the optical emission likely originates from a single emission process, synchrotron emission from the forward shock (i.e. emission from the shocked ambient medium), in which ordered fields are much less likely to be present$^{5}$.  
 This then allows measurements of the geometry of the jet through monitoring of the late-time polarization angle$^{2,3,4,16,17}$, but crucially also offers a 
simple test for afterglow micro-physics via circular polarimetry -  detailed models have been developed, but observational 
attempts have largely focussed on radio reverse shocks, producing only upper limits$^{18}$, and a single optical non-detection$^{17}$.

GRB\,121024A was detected by the Burst Alert Telescope (BAT) onboard the {\it Swift} satellite at 02:56:12 UT on 24 October 2012 $^{19}$; a redshift of  $z = 2.298$ was found shortly after (Methods). We obtained imaging polarimetry observations with the FOcal Reducer and low dispersion Spectrograph (FORS2) on the Very Large Telescope (VLT), using a Wollaston prism and quarter and half wavelength plates (Extended Data Fig. 1 and 2). Observations with the $R_{\rm special}$ filter commenced at 2.57 hours after the burst, when the afterglow had a brightness of $R\sim19.8$ mag. After two sets of linear polarimetry ($P_{\rm lin}$), we obtained four consecutive circular polarization 
($P_{\rm cir}$) measurements, followed by a further 9  $P_{\rm lin}$ measurements (of which 4 on the second night). Data reduction and calibration follow standard procedures$^{17}$  (Methods). Simultaneous with the polarimetry, we monitored the afterglow optical light curve with the Gamma-Ray Burst Optical/Near-Infrared Detector (GROND) instrument (Methods). In the following we will use a notation where the flux density $F$ depends on frequency and time as $F\propto t^{-\alpha}\nu^{-\beta}$, with temporal decay index $\alpha$ and spectral energy index $\beta$. 

The X-ray light curve obtained by {\it Swift} (retrieved from the online {\it Swift}/XRT GRB lightcurve repository$^{20}$) is well described by three power law segments, where the first break occurs at $t_{\rm break,1} = 619 ^{+199}_{-348}$ s and a second break at $t_{\rm break,2} = 3.4^{+1.5}_{-2.2} \times 10^4$ s (errors at 90\% confidence level). In the GROND light curve we find evidence of a break at a time consistent with this last X-ray break (Extended Data Fig. 3). A combined fit to the X-ray and GROND data, using a smoothly broken power law and a host galaxy component, gives best fitting break 
time $t_{\rm break,2} = 3.72 \pm 0.07 \times10^4$ seconds (see Methods), pre-break light curve decay indices $\alpha_{\rm pre,opt} = 0.93 \pm 0.02$, 
$\alpha_{\rm pre,X-ray} = 0.96 \pm 0.11$; post-break light curve decay indices $\alpha_{\rm post,opt} = 1.25 \pm 0.04$, $\alpha_{\rm post,X-ray} = 1.67 \pm 0.10$ (uncertainties 
are 1$\sigma$). The occurrence of a light curve break simultaneously in X-rays and optical wavelengths is suggestive of a jet break origin. The X-ray+GROND spectral energy distribution is best fit with a single power law with $\beta = 0.88 \pm 0.01$ and an optical extinction $A_V = 0.22 \pm 0.02$ magnitudes (Methods; Extended Data Fig. 4). The pre-break temporal and spectral indices agree with the standard fireball closure relations$^{21}$ in the situation where the synchrotron cooling 
frequency $\nu_{c}$ and peak frequency $\nu_m$ are below both optical and X-ray frequencies$^{21}$. 

Figure 1 shows the observed $P_{\rm lin}$ behaviour. Initially the source starts out $\sim5\%$ polarized (significantly lower than the $\sim70\%$ expected for a perfectly ordered magnetic field), which subsequently decays to lower levels, while the polarization angle is remarkably constant. This shows that the magnetic field directions are largely random, i.e., the coherence scales of the field in the blast wave are small, but their directions are confined to the plane of the shock (the detected polarization is attributed to a somewhat off-axis viewing angle). In the second night of data, the polarization angle is markedly different, consistent with a $90^{\circ}$ angle change (Figure 1). We consider this an unambiguous detection of the 90$^{\circ}$ angle change predicted to occur around the jet break time of a homogeneous, not sideways spreading jet$^{4}$. The exact time at which the angle change occurs is dictated by viewing angle$^{4}$. The observed angle change shows that any ordered magnetic fields in the forward shock are weak
if at all present$^{22}$.

We acquired the $P_{\rm cir}$ measurements shown in Figure 2 between the second and third $P_{\rm lin}$ datapoints. Under the assumption that during this interval no variability in $P_{\rm cir}$ is expected (the time covered is small compared to the time after burst), we combine the 4 measurements together, and measure 
$P_{\rm cir} = 0.61 \pm 0.13\%$ (see Methods). We estimate the linear polarization degree during the circular polarimetry interval to be $P_{\rm lin} \sim 4\%$: the afterglow shows a ratio $P_{\rm cir} / P_{\rm lin} \sim 0.15$, several orders of magnitude above basic model predictions ($\sim10^{-4}$ at optical wavelengths$^{6,7,8}$, see Methods), and other relativistic jet sources$^{23}$ (Figure 3).

A high level of $P_{\rm cir}$ can be completely intrinsic to the source (e.g. the $P_{\rm cir}$ of the synchrotron emission from the source) or have its origin in propagation effects within the source (e.g. Faraday conversion $P_{\rm lin} \rightarrow P_{\rm cir}$, which is effective in a hot, relativistic medium) or by dust scattering effects along the line of sight. Plasma propagation effects have been shown to be strong at long wavelengths (close to the synchrotron self absorption frequency $\nu_a$), but negligible at optical wavelengths$^{6,7,8}$: it 
is very unlikely these effects play a role here. 

The influence of dust is limited to the host galaxy, as the Galactic extinction towards GRB\,121024A is very small, $E(B-V) = 0.10$ (Methods). 
Dust affects optical $P_{\rm cir}$ through four possible routes: multiple scattering in an optically thick medium of dust grains; dichroic scattering by (somewhat aligned) non-spherical dust grains; dichroic extinction of linearly polarized radiation by (somewhat aligned) non-spherical dust grains; and the scattering of linearly polarized radiation by randomly oriented dust particles (if the polarization is not in, or perpendicular to, the scattering plane). The expected $P_{\rm cir}$ through the latter three effects depends strongly on the degree of alignment, source inclination and the total amount of dust involved, for which line of sight extinction may be a proxy$^{24,25}$. The weak line of sight host galaxy extinction, $A_V = 0.22 \pm 0.02$
 magnitudes (see Methods), argues against multiple-scattering effects $^{25}$. To rule out the other possibilities, we simulated the efficiency of conversion of $P_{\rm lin} \rightarrow P_{\rm cir}$ through single dust scattering (that is, where each photon has been scattered at most once by a dust particle), finding maximal conversion values  at high and low scattering angle regimes of $\sim10\%$ at  $\sim100^{\circ}$,
 and $\sim8\%$ at $\sim20^{\circ}$, respectively. Models based on partially aligned dust grains have lower efficiency. If a large part of $P_{\rm cir}$ is in fact due to dust, we expect a large fraction, if not all, of the $P_{\rm lin}$ to be caused by dust scattering too. The linear polarization curve (Figure 1) shows no signs of this: we would not expect to see a constant angle pre-break, nor for $P_{\rm lin}$ to reach near zero at any point, nor a clear 90$^{\circ}$ angle change over the jet break$^{26}$. In addition, light scattered at large scattering angles has not had time to reach us yet (dust very close to the GRB gets destroyed by the GRB prompt emission). After eliminating plasma propagation effects and dust scattering, we therefore conclude that the measured $P_{\rm cir}$ is likely largely or fully intrinsic to the afterglow. It is of some interest to compare this with the only other afterglow with deep measurements, GRB\,091018, which showed limits$^{17}$: $P_{\rm cir} < 0.15\%$ (2$\sigma$), and $P_{\rm cir} / P_{\rm lin} <1$ (Figures 1, 2, 3).
 
The origin of the optical circular polarization in the afterglow of GRB\,121024A is puzzling and unexpected. We expect polarization from the external shock synchrotron emission of  $P_{\rm cir}
\simeq \gamma_{e}^{-1}$, where $\gamma_e$ is the random Lorentz factor of the electrons emitting the optical radiation -- under the assumption of isotropic electron pitch angle 
distribution and perfectly ordered magnetic field$^{8}$ (see Methods). In this situation, the observed $P_{\rm cir}$ would imply extremely low values of 
$\gamma_e$. Furthermore, since $P_{\rm cir}$ and $P_{\rm lin}$ are both expected to be reduced from the value for a perfectly ordered field case by the field randomness in the 
same way, the expected $P_{\rm cir} / P_{\rm lin}$ ratio is the same as for a perfectly ordered field, again scaling with $\gamma_{e}^{-1}$ in the case of an isotropic electron 
distribution$^{8}$. Therefore, the observed polarimetric behaviour poses a challenge to the long-standing assumption of isotropic electron pitch angle distributions in the GRB 
forward shock afterglow. Pitch angle anisotropy has been postulated before as a possible explanation of GRB phenomena as varied as steep decay phases in X-ray light curves$^{27}$ and spectra of prompt emission$^{28}$, but evidence for this has been lacking. Further hints towards a more complicated structure of emission and acceleration regions come from observations as varied as high-energy emission in GRBs$^{29}$ and fast variability of high energy emission in quasars$^{30}$. 
The circular polarization of GRB afterglows as well as quasars (see Fig. 3) offer a new line of evidence required to guide theoretical studies$^{9,10}$.

\noindent \underline{Methods Summary}\\
Both linear and circular polarimetry of the afterglow of GRB\,121024A was performed using the FORS2 instrument on the VLT, using the $R_{\rm special}$ filter in 
imaging polarimetry mode, utilising a Wollaston prism and half- and quarter wavelength plates. We used four plate angles for the linear polarimetry and two for 
circular polarimetry, in order be able to use beam-switching to reduce systematic errors, and used aperture photometry to measure source fluxes. 
Measurements from polarimetric sequences taken on the second night after burst were combined together to increase signal to noise. We used field stars to measure the
linear polarization induced by Galactic dust, fitting the distribution in Stokes $Q,U$ with a two dimensional Gaussian distribution. 
We monitored the optical afterglow brightness in seven photometric filters using the GROND instrument, and fit the X-ray data from the {\it Swift} satellite together with the
GROND data to establish the presence of a late-time break in the light curve, and fit the X-ray to optical spectral energy distribution of the afterglow. Finally, we demonstrate how the observed high level of circular polarization contradicts theoretical estimates, and how anisotropy in the electron pitch angle distribution may explain the observed
ratio of optical circular to linear polarization.

\noindent 
\noindent  [1] Piran, T. Magnetic fields in Gamma-ray bursts: a short overview. {\it AIP Conference 
    Proceedings} {\bf 784}, 164-174 (2005)

\noindent [2] Sari, R. Linear polarization and proper motion in the afterglow of beamed gamma-ray 
  bursts. {\it Astrophys. J}. {\bf 524}, L43-L46 (1999) 

\noindent [3] Ghisellini, G. \& Lazzati, D.  Polarization light curves and position angle variation of 
  beamed gamma-ray bursts. {\it Mon. Not. R. Astron. Soc}. {\bf 309}, L7-L11 (1999)

\noindent  [4] Rossi, E. M., Lazzati, D., Salmonson, J. D., Ghisellini, G. The polarization of afterglow   
      emission reveals gamma-ray bursts jet structure. {\it Mon. Not. R. Astron. Soc}. {\bf  354},   
      86-100 (2004)

 \noindent [5] Mundell, C.~G. {\it et al}. Highly polarized light from stable ordered magnetic fields in 
      GRB\,120308A. {\it Nature} {\bf 504}, 119-121 (2013) 

\noindent  [6] Matsumiya, M. \& Ioka, K. Circular polarization from gamma-ray burst afterglows.  
   {\it  Astrophys. J}.  {\bf 595}, L25-L28 (2003)

\noindent  [7] Sagiv, A., Waxman, E. \& Loeb, A. Probing the magnetic field structure in gamma-ray   
    bursts through dispersive plasma effects on the afterglow polarization. {\it Astrophys. J}.  {\bf 615}, 
    366-377 (2004)

\noindent [8] Toma, K., Ioka, K. \& K., Nakamura, T. Probing the efficiency of electron-proton coupling in 
    relativistic collisionless shocks through the radio polarimetry of gamma-ray burst
    afterglows. {\it Astrophys. J}. {\bf 673}, L123-L126 (2008)

\noindent [9] Spitkovsky, A. Particle Acceleration in Relativistic Collisionless Shocks: Fermi Process 
      at Last? {\it Astrophys. J}.  {\bf 682}, L5-L8 (2008) 

\noindent [10] Spitkovsky, A. On the Structure of Relativistic Collisionless Shocks in Electron-Ion Plasmas. 
     {\it Astrophys. J}.  {\bf 673}, L39-L42 (2008) 

\noindent  [11] Hededal, C. B. \& Nishikawa, K.-I. The influence of an ambient magnetic field on 
   relativistic collisionless plasma shocks.  {\it Astrophys. J}. {\bf 623}, L89-L92 (2005)

\noindent  [12] Pudritz, R. E., Hardcastle, M. J.,  Gabuzda, D. C. Magnetic fields in astrophysical jets: 
    from launch to termination. {\it Space Science Reviews}, {\bf 169}, 27-72 (2012)

\noindent [13] Lyutikov, M. Magnetocentrifugal launching of jets from discs around Kerr black holes. 
    {\it Mon. Not. R. Astron. Soc} {\bf 396},  1545-1552 (2009)

\noindent [14] G\"{o}tz, D., Laurent, P., Lebrun, F.,  Daigne, F., Bosnjak, Z. Variable polarization measured 
   in the prompt emission of GRB 041219A using IBIS on board INTEGRAL. {\it Astrophys. J}.  
   {\bf 695}, L208-L212 (2009)

\noindent [15] Yonetoku, D. {\it et al}. Magnetic structures in gamma-ray burst jets probed by gamma-ray 
  polarization.  {\it Astrophys. J}. {\bf 758}, L1-L6 (2012)

\noindent [16] Greiner, J. {\it et al}. Evolution of the polarization of the optical afterglow of the gamma-ray 
  burst GRB 030329. {\it Nature} {\bf 426}, 157-159 (2003)

\noindent [17] Wiersema, K. {\it et al}.  Detailed optical and near-infrared polarimetry, spectroscopy and  
  broad-band photometry of the afterglow of GRB\,091018: polarization evolution. {\it Mon. 
  Not.   R. Astron. Soc}. {\bf 426}, 2-22 (2012)

\noindent [18] Granot, J. \& Taylor, G.ÊB.  Radio flares and the magnetic field structure in gamma-ray 
  burst outflows.  {\it Astrophys. J}.  {\bf 625}, 263-270 (2005)

\noindent [19] Pagani, C. {\it et al}. GRB\,121024A: Swift detection of a burst with an optical counterpart.
  {\it GCN Circ}. {\bf 13886} (2012)

\noindent [20] Evans, P. A. {\it et al}.  Methods and results of an automatic analysis of a complete sample 
  of Swift-XRT observations of GRBs.  {\it Mon. Not. R. Astron. Soc}. {\bf 397},  1177-1201 
  (2009)

\noindent [21] Sari, R., Piran, T., Narayan, R. Spectra and lightcurves of gamma-ray burst afterglows. 
  {\it Astrophys. J}. {\bf 497}, L17-L20 (1998) 

\noindent [22] Granot, J. \& K\"{o}nigl, A. Linear polarization in gamma-ray bursts: the case for an 
  ordered magnetic field.  {\it Astrophys. J}. {\bf 594}, L83-L87 (2003)

\noindent [23] Hutsem\'{e}kers, D., Borguet, B., Sluse, D., Cabanac, R., Lamy, H. Optical circular 
  polarization in quasars. {\it Astron. Astrophys}. {\bf 520}, L7 (2010)

\noindent [24] Whitney, B. A. \& Wolff, M. J.  Scattering and Absorption by Aligned Grains in 
  Circumstellar Environments. {\it Astrophys. J}. {\bf 574}, 205-231 (2002)

\noindent [25] Fukue, T. {\it et al}.  Near-Infrared Circular Polarimetry and Correlation Diagrams in the 
  Orion Becklin-Neugebauer/Kleinman-Low Region: Contribution of Dichroic Extinction.
  {\it Astrophys. J}. {\bf 692}, L88-L91 (2009)    

\noindent [26] Lazzati, D. {\it et al}. Intrinsic and dust-induced polarization in gamma-ray burst afterglows: 
  The case of GRB 021004. {\it Astron. Astrophys}. {\bf 410}, 823-831 (2003)

\noindent [27] Beloborodov, A. M., Daigne, F., Mochkovitch, R., Uhm, Z. L. Is gamma-ray burst  
      afterglow emission intrinsically anisotropic? {\it Mon. Not. R. Astron. Soc}. {\bf 410}, 2422-2427 
      (2011)

\noindent [28] Lloyd-Ronning, N. M. \& Petrosian, V. Interpreting the behavior of time-resolved 
 gamma-ray burst spectra.  {\it Astrophys. J}.  {\bf 565}, 182-194 (2002).

\noindent [29] Kouveliotou,  C.  {\it et al}.  NuSTAR Observations of GRB 130427A Establish a Single Component Synchrotron Afterglow Origin for the Late Optical to Multi-GeV Emission.  {\it Astrophys. J}.  {\bf 779}, L1 (2013)

\noindent [30] Ghisellini, G., Tavecchio, F., Bodo, G.,  Celotti, A. TeV variability in blazars: how fast can it be? {\it Mon. Not. R. Astron. Soc}. {\bf 393}, L16-L20

\newpage


\begin{figure}[h!]
\includegraphics[width=12cm]{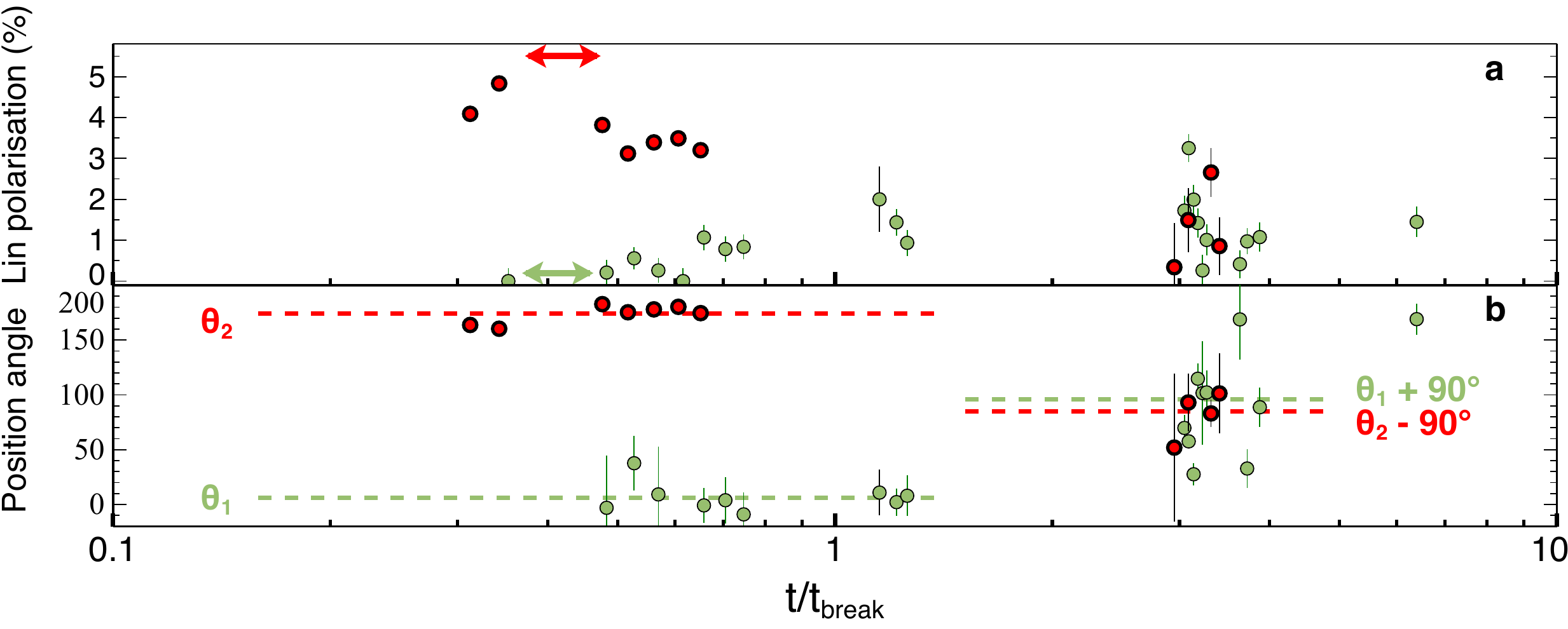}
\caption*{ {\bf Figure 1. Linear polarization of the afterglow of GRB 121024A}. {\bf a},  
The linear polarization degree as a function of time; {\bf b},  the polarization angle is shown.
Red points are GRB\,121024A,  green points are for the only other {\it Swift} GRB
afterglow with extensive polarimetry, GRB\,091018$\,^{17}$ (error bars  are  $1\sigma$).
 The horizontal axis marks the time since trigger, normalised by the jet break time ($t_{\rm break,2} = 3.72 \pm 0.07 \times10^4$ s). 
 $\theta_1$ and $\theta_2$ show the average pre-jet break angles;  $\theta_{1,2}+90$ demonstrate the 90$^{\circ}$ angle change predicted for jet breaks of uniform jets.  Horizontal bars show the timespan over which the circular polarimetry was obtained.
}
\end{figure}

\begin{figure}[h!]
\includegraphics[width=8cm]{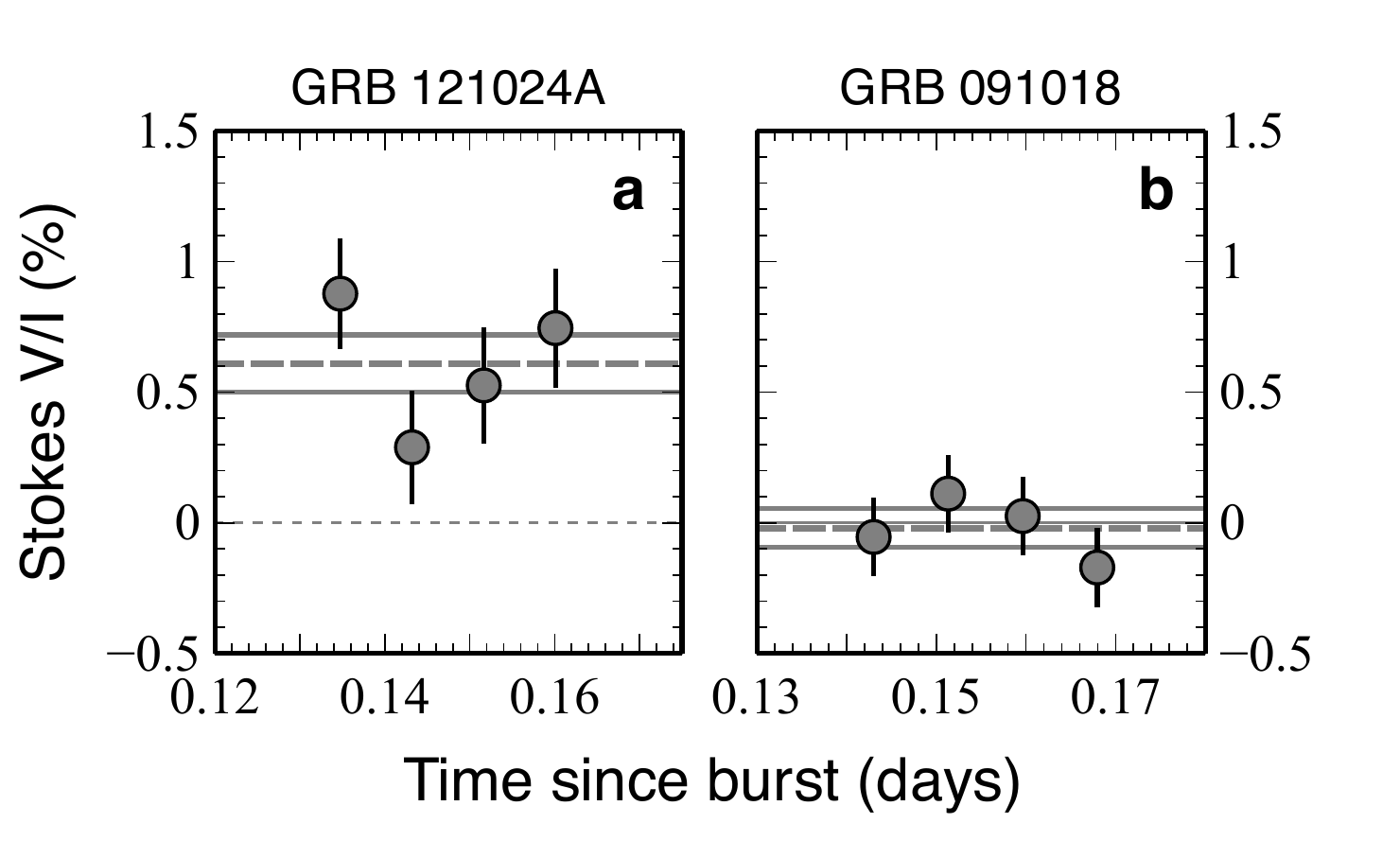}
\caption*{{\bf Figure 2. Optical circular polarization measurements of the afterglow of GRB 121024A}. {\bf a}, Optical circular polarimetry of the 
afterglow GRB\,121024A; {\bf b}, Optical circular polarimetry for GRB\,091018$^{17}$ (error bars are $1\sigma$).  The horizontal axis 
shows the time since BAT trigger in the observer frame, on the vertical axis is the polarization as $V/I$, in percent. The dotted  line indicates $V/I = 0$, the dashed  line 
shows the measured $V/I$ from the combined data points, and $1\sigma$ uncertainty values around the best combined data value are shown with solid lines.   
}
\end{figure}
\begin{figure}[h!]
\includegraphics[width=10cm]{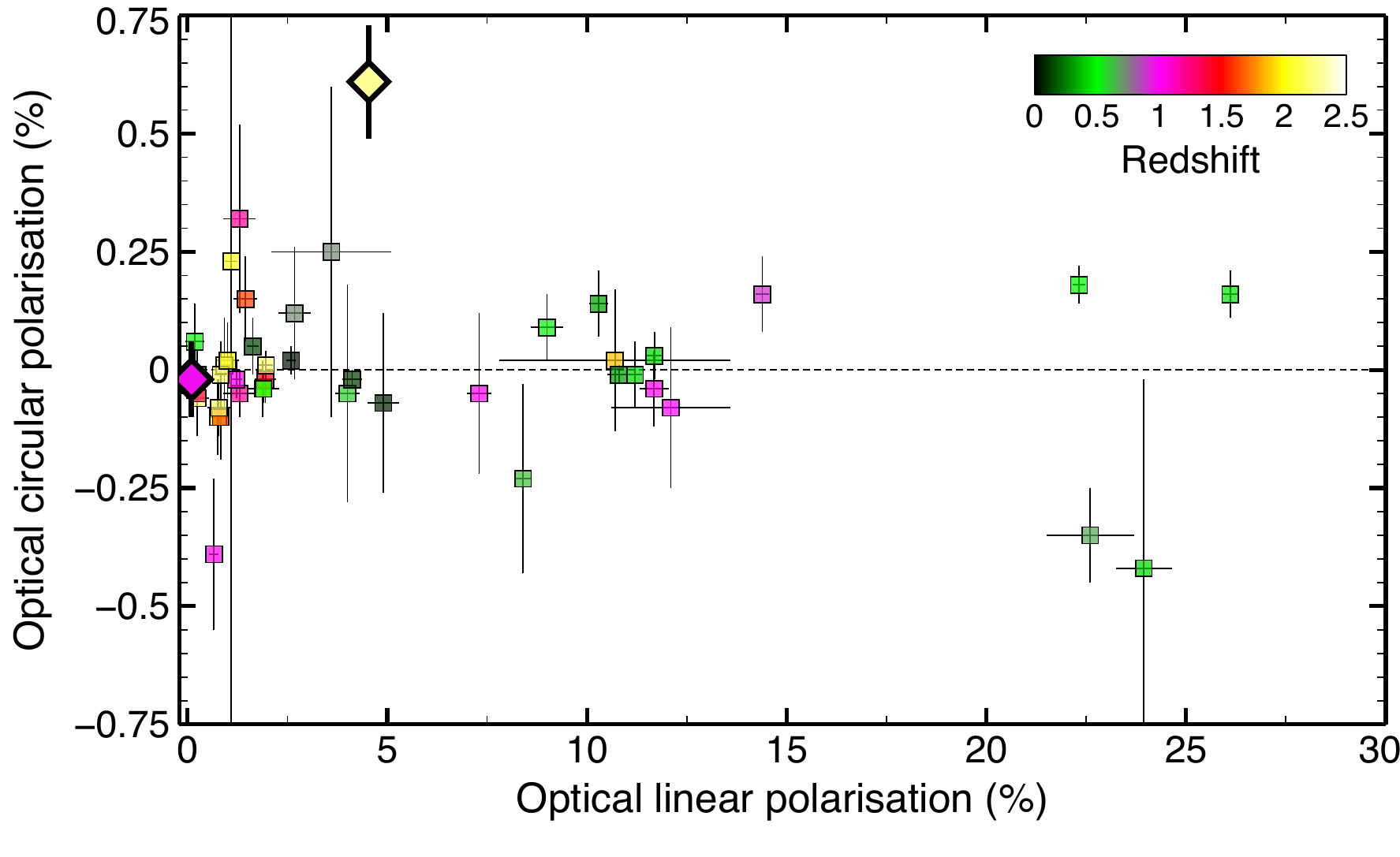}
\caption*{ {\bf Figure 3. Optical polarimetry of quasars and GRB afterglows compared}.  
polarization measurements for GRBs 121024A and 091018 (diamond symbols) are shown with optical polarimetry of quasars$^{23}$ (squares); error bars are $1\sigma$. The data points are colour-coded for redshift. The few quasars with $>3\sigma$ detection of optical circular polarization have high values ($\geq20\%$) of  linear polarization. 
GRB\,121024A has the highest detected level of optical circular polarization, and at $z = 2.298$ is the highest redshift object with a 
detection, showing the potential for GRBs as probes of possible cosmological propagation effects. We caution that the range of redshifts means that different restframe wavelengths are compared.
}
\end{figure}

\include{supplementaryinfo}

\noindent \underline{Methods}

\noindent {\bf Linear and circular polarimetry}\\
GRB 121024A triggered the Burst Alert Telescope (BAT) onboard the Swift satellite at 02:56:12 UT on 24 October 2012 (Swift trigger 536580)$^{19}$. We will use this trigger time as $t_0$ (i.e. time since burst of an observation is $t = t_{\rm obs} - t_0$) throughout. The prompt emission shows the burst likely belongs to the class of long bursts, with a duration $t_{\rm 90} = 69 \pm 32$ seconds$^{31}$. An X-ray and optical afterglow was found by the Swift X-ray telescope (XRT) and UV-optical telescope (UVOT)$^{19}$. The redshift, $z = 2.298$, was found through afterglow spectroscopy with the X-shooter instrument on the Very Large Telescope$^{32}$. Based on the initial brightness of the UVOT afterglow$^{19}$ we activated our VLT polarimetry programme (programme 090.D-0789, PI Wiersema).

The data acquisition strategy, reduction and analysis closely follows that of our recent paper on GRB 091018$^{17}$, which in turn follows the specific recommendations set out for FORS polarimetry$^{33}$. We summarise those methods here, and give additional details specific to the case of GRB\,121024A. 

Observations in the FORS2 $R_{\rm special}$ filter started at $9.2 \times 10^3$ sec after burst. Following an acquisition image (Extended Data Figure 1), polarimetry was acquired in imaging polarimetry (IPOL) mode. In this instrument mode, a super-achromatic half- or quarter wavelength plate is used (for linear and circular polarimetry, respectively), after which a Wollaston prism is used to split
the light into two beams, the so-called ordinary and extraordinary beam (hereafter the {\it o} and {\it e} beam) that have perpendicular polarization. These beams are imaged simultaneously: a slit mask is used to prevent overlap of the two beams on the chip (Extended Data Figure 2). All linear polarization measurements are obtained using four rotation angles (0, 22.5, 45 and 67.5 degrees) of the half wavelength plate; for the circular polarimetry we used two angles (-45 and +45 degrees) of the quarter wavelength plate. The GRB afterglow was positioned in the middle of the mask opening at the default (on-axis) position for FORS2 observations, on chip 1. 
A small number of linear polarization measurements were taken with a small dither ($\sim16$ pixels in Y direction) to eliminate any potential effects of bad pixels / columns; none of these problems were apparent. Extended Data Tables 1 and 2 list the polarimetric observations and their resulting measurements.

Data reduction was done within {\sc IRAF}, using bias and sky flat frames obtained on the same 
nights. Analysis of the data was done via aperture photometry, using IRAF scripts developed for this purpose, using the following method. We perform aperture photometry of all point sources present on both FORS2 chip 1 and 2, on the {\it o} and {\it e} images, using an aperture radius of 1.5 times the on-frame full width at half maximum (FWHM) of the point spread function (PSF). The PSF is found via a Gaussian fit on all point source objects, on a per image 
basis, and is determined independently for the  {\it o} and  {\it e} beam as small differences in PSF shape may occur between the beams, particularly for objects far off-axis. The PSF value was used to set the aperture size for e.g. the  {\it o} beam on chip 1 is the weighted average of the FWHM of all point sources in that beam, chip and image. The sky subtraction was done using an annulus of inner and outer radii 3 and 4 times the FWHM, respectively. We only included sources for which the sky annulus was fully contained within the mask and showed no signs of saturation.
Using this procedure we measure fluxes $f_o$ and $f_e$ for all point-like objects in all frames. Errors are determined using 
$\sigma^2_{\rm source} = g^{-1} \times f + (n_{\rm A} + n_{\rm A}^{2} / n_{\rm B}) \sigma^2_{\rm B} / {\rm pixel}$ , where $g$ is the gain, $f$  the 
flux ($f_o$ and $f_e$ , source minus the mean background) in the aperture, $n_{\rm A}$ and $n_{\rm B}$ the number of pixels in aperture and background region; and 
$\sigma_{\rm B/pixel}$ the variance per pixel in the sky level within the annulus. 

The read noise contribution to the errors is negligible in these data. Images are not combined (i.e. the depth of coverage per pixel is 1). The purpose of using the four angles for linear polarization, and two for the circular, is that the beams switch, eliminating several systematic errors (e.g. flat field defects), resulting in increased reliability$^{33}$. We use the fluxes to derive the normalised flux difference at angle number $i$ as $F_i = \left(f_{o,i} - f_{e,i}\right) / \left(f_{o,i} + f_{e,i}\right) = \left(f_{o,i} - f_{e,i}\right) / I$. We use the Stokes parameters $(U,Q,V,I)$ to describe the source polarization state, often in normalised form $(U/I, Q/I, V/I)$. These can be expressed in terms of $F_i$ as: 
$Q/I = \frac{2}{N} \sum\limits_{i=0}^{N-1} F_{i} \cos(i\pi/2)$;  and $U/I =  \frac{2}{N} \sum\limits_{i=0}^{N-1} F_{i}\sin(i\pi/2)$, 
where $N$ is the number of half wavelength plate positions (four positions in this case)$^{17,33}$. The circular polarization Stokes parameter $V/I$ is similarly computed 
as $V/I = \frac{1}{2} \left(F_{45} - F_{-45}\right)$. 
The benefit of the multiple angle observations is clear from these equations. Further increasing the accuracy by observing at even more angles (e.g. eight angles for linear polarization) is difficult for sources as faint as the afterglow of GRB\,121024A - it would lead to observing times for single data points that are close to the polarization variability timescale. 

We now proceed to correct for the linear polarization induced by dust in our own Galaxy which, along the line of sight to GRB\,121024A, 
has $E(B-V) = 0.10\,^{34}$. Using the relation $P_{\rm lin} \leq 0.09 \times E(B-V)\,^{35}$ this corresponds to a maximum induced polarization 
of $\sim$0.9\%, which may be a significant fraction of the detected polarization. To correct for Galactic dust induced polarization, we use the same methods as we used for 
GRB 091018$^{17}$: we fit the field star $Q,U$ distribution with a two dimensional Gaussian function. We use three cuts on the sources entering the distribution: we require a polarimetric error $<0.9$\%, a polarization value below 1.5\% and a radial distance to the GRB position smaller than 2.5 arc minutes. This last cut is to counter the effects of instrumental polarization: the FORS2 instrument shows a broadly radial instrumental linear polarization pattern, with nil polarization on-axis (where the GRB is positioned), and values increasing with radial distance$^{33,36,37}$. By picking this cut-off we prevent the instrumental pattern from influencing the field objects' $Q,U$ centroid determination too much, while still retaining enough sources for a reliable fit. We note that the number of bright field point sources within the mask is low in the case of 121024A (87 data points from 9 individual stars enter the fit, considerably lower than in the case of 091018\,$^{17}$), resulting in a somewhat larger uncertainty in the Galactic dust induced $Q,U$ value. We find a value of   
$Q_{\rm Gal} = -0.0020, U_{\rm Gal} = -0.0046$. We use the standard deviation of the fitted 2D Gaussian as a measure for the uncertainty on the Galactic dust induced Stokes parameters, finding $\sigma_{Q_{\rm Gal}} = 0.0040, \sigma_{U_{\rm Gal}} = 0.0043$. We correct the afterglow $Q,U$ values by subtracting the Galactic dust induced $Q,U$ values. To facilitate comparison with models, we use the Stokes parameters to express the linear polarization in terms of the polarization degree $P_{\rm lin}$ and polarization angle $\theta$ as $P_{lin} = \sqrt{\left(Q^{2} + U^{2}\right)} / I$ and $\theta = \frac{1}{2} \arctan(U/Q)$, where coordinates are chosen such that $\theta = 0$ degrees for North, $\theta = 90$ degrees for East. In the conversion from $Q,U$ to $P_{\rm lin}, \theta$ we account for the effects of polarization bias$^{38,39}$ in the same manner as done for the data of GRB 091018$^{17,33}$. We correct the angles $\theta$ for the FORS2 instrumental zero angle offset$^{37}$. Similarly we can compute $P_{\rm cir} = \sqrt{(V/I)^{2}}$: the sign of $V/I$ gives the polarization direction (clockwise or counter-clockwise). We note that both the instrument induced circular polarization and linear to circular polarization crosstalk are well studied for the FORS instruments and far below our detection levels on the optical axis (where the afterglow is positioned)$^{33,36,37}$, as such we expect no instrumental contribution to the detected $V/I$. A small amount of instrumental circular polarization is expected to be present for sources observed far off axis$^{36,37}$, far from the GRB position. Since there are no bright (but unsaturated) stars near the GRB position, it is therefore not possible to use field sources as secure independent secondary standards$^{33,36,37}$. The circular polarization induced by scattering by the dust in our own Galaxy, with $E(B-V) = 0.10$ and the induced linear polarization values as above, does not contribute significantly to the observed $P_{\rm cir}$\,$^{40}$.

In the second night data, the afterglow of GRB\,121024A shows a low level of linear polarization, which means that the uncertainty on the angle $\theta$ is relatively large, 
as $\sigma_{\theta} = \sigma_{P_{\rm lin}} /  2 P_{\rm lin}$. Nevertheless a clear difference in angle is visible with respect to the first night of data, consistent with a 90 degree  change in polarization angle: the angle changes from $172 \pm 2$ degrees  to $85 \pm 10$ degrees (Extended Data Table 1). The two first observations have angles somewhat discrepant from the pre-jet break average (Extended Data Table 1).
Unfortunately we were not able to acquire infrared polarimetry simultaneous with the the $R$ band polarimetry for scheduling reasons - simultaneous multi wavelength polarimetry is the best way to directly measure polarization induced by scattering off dust particles in the host galaxy$^{34,40}$. However, the measured line of sight extinction and indeed the linear polarization light curve itself, provide sufficient evidence that the host galaxy dust induced linear polarization must be small: the measured $A_V \sim 0.22$ mag gives a limit on the induced linear polarization of $<0.7\%$ (assuming a Milky Way like polarization curve$^{35}$).


\noindent {\bf Lightcurves and spectral energy distribution}\\
\noindent The {\it Swift} XRT light curve is well described by three power law segments, with parameters (errors at 90\% confidence level) $\alpha_1 = 1.78^{+0.49}_{-0.10}$, 
$t_{\rm break,1} = 619^{+199}_{-348}$ s, $\alpha_2 = 0.83^{+0.13}_{-0.22}$, $t_{\rm break,2} = 3.4^{+1.5}_{-2.2} \times 10^4$ s and 
$\alpha_3 = 1.70^{+0.27}_{-0.36}$\,$^{20}$. The uncertainty in the late break time is relatively large, as there are few data points post-break.

The optical afterglow of GRB 121024A was observed extensively with the seven channel 
($g', r', i', z', J, H, Ks$ filters) Gamma-Ray Burst Optical and Near-Infrared Detector (GROND$^{42}$) instrument, mounted on the 
ESO 2.2m telescope at La Silla, Chile, providing photometric monitoring simultaneous to the polarimetry (Extended Data Table 3). These GROND observations will be described in more detail in a forthcoming, separate publication, but below we give some properties that are required to interpret the polarimetry.
 In addition to the GROND data, we observed the position of GRB 121024A with the ACAM instrument on the William Herschel Telescope (WHT) on 15 February 2013 (i.e. 114.8 days after burst), with the aim of fixing the flux contribution of the host galaxy to the late time light curve. We acquired $14 \times 180$ seconds exposures in Sloan $r$ filter in relatively poor seeing conditions ($\sim1.8$''). The resulting image (Extended Data Figure 1) shows a clear detection of a source at the afterglow position, which we identify as the host galaxy of this GRB, confirming that a late time detection in GROND data in $g, r, i$ is dominated by the host. Its brightness, $r' = 24.03 \pm 0.20$, places the host at the bright end of the host galaxy luminosity distribution at this redshift$^{43}$. Taking this bright host galaxy magnitude into account, a light curve break at around the same time as the break in XRT data is apparent (Extended Data Fig. 3).  
We perform a joint fit to the XRT and GROND light curves (where only XRT data after the first XRT light curve break, $t_{\rm break,1}$ are used) to constrain the break time.  
We fit using a model that consists of a smoothly broken power law, generally defined as
\[
F_{\nu}(t) = F_{\nu}(t_{\rm break})\left(\left(\frac{t}{t_{\rm break}}\right)^{\alpha_{1}s} + \left(\frac{t}{t_{\rm break}}\right)^{\alpha_{2}s} \right)^{-1/s},
\]
where $t_{\rm break}$ is the break time, and $\alpha_{1}$ and $\alpha_{2}$ are pre- and post-break lightcurve indices, and $s$ is the break smoothness parameter. 
The pre- and post break slopes and break smoothness are free parameters in our fit, and the break time is fixed to be the same for X-ray and optical/infrared wavelengths (i.e. an achromatic break, and in addition a host galaxy contribution to the optical and  
infrared fluxes. This results in an acceptable fit statistic (reduced $\chi^2 = 157.38/132 = 1.192$), and the following parameters (uncertainties are 1$\sigma$):  break 
time $t_{\rm break,2} = 3.72 \pm 0.07 \times 10^4$ s and break smoothness $s = 5.01 \pm 0.01$; pre-break light curve decay indices $\alpha_{\rm pre,opt} = 0.93 \pm 0.02$, 
$\alpha_{\rm pre,X-ray} = 0.96 \pm 0.11$; post-break light curve decay indices $\alpha_{\rm post,opt} = 1.25 \pm 0.04$, $\alpha_{\rm post,X-ray} = 1.67 \pm 0.10$. 
We identify this late, achromatic, break $t_{\rm break,2}$ with a so-called jet break.
The resulting fit is shown in Extended Data Figure 3. While in each optical band  there are only a few data points post-break, GROND observes in seven bands simultaneously, making the break  significant. The relatively shallow post-break optical decay is likely caused by the combination of bright host and smooth break: by the time the light curve asymptotes to its post-break index it is dominated by host galaxy light.

The line of sight extinction in the host galaxy and the spectral slopes are found by fitting an XRT+GROND spectral energy distribution (at time 11085 s after trigger) with a SMC-like extinction law$^{44}$. The best fit, with reduced $\chi^2 = 1.04$, is obtained by a single power law (a broken power law does not result in significant fit improvement) with parameters  $\beta = 0.88  \pm 0.01$, and a optical extinction in the $V$  band of $A_V = 0.22 \pm 0.02$ magnitudes. 

The fact that X-ray and optical/infrared wavelengths have the same spectral index and that the pre-break decay indices are (within errors) identical, suggests that X-ray and optical are both in the same spectral regime, likely $\nu > \nu_{\rm c}$. The achromatic nature of the light curve break is consistent with a jet break interpretation, supporting our interpretation of the linear polarization behaviour of this afterglow.

\noindent {\bf Circular polarization of synchrotron emission}\\
\noindent The linear and circular polarization degrees of the optically-thin synchrotron emission from the electrons with a spectrum given by 
$N(\gamma_{e},\alpha) = K \gamma_{e}^{-p} f(\alpha)$, where $\gamma_{e}$ is the electron Lorentz factor, $\alpha$ is the electron pitch angle, and $K$ is the normalisation factor, are given by$^{45,11}$ :
\[ P_{\rm lin} = \frac{p + 1}{p + \frac{7}{3}},
\]
\[
P_{\rm cir} = \frac{1}{\gamma_e}\frac{(2+p)\cot\theta + g(\theta)}{p}\frac{p+1}{p + \frac{7}{3}}\frac{\Gamma\left(\frac{3p + 8}{12}\right)\Gamma\left(\frac{3p + 4}{12}\right)}{\Gamma\left(\frac{3p + 7}{12}\right)\Gamma\left(\frac{3p - 1}{12}\right)}
\]
Here $\theta$ is the viewing angle with respect to the magnetic field direction, and we have defined
\[
\left. g(\theta) \equiv \frac{1}{f(\theta)}\frac{df(\alpha)}{d\alpha}\right\vert_{\alpha=0}
\]
These two equations are valid when $g(\theta) \ll \gamma_e$.

If the electron pitch angle distribution is isotropic, i.e.  $g(\theta) = 0$, then $P_{\rm cir} \sim \gamma_e^{-1}$.  
This simply means that the circular polarization contributions of electrons with pitch angles $\theta+\alpha$ and $\theta-\alpha$ nearly cancel out, and the remaining $P_{\rm cir}$ scales with the angular size of the beaming cone of the synchrotron emission, $\gamma_e^{-1}$.  

The electrons with Lorentz factor $\gamma_e$ mainly contribute to the synchrotron emission at frequency  
$\nu = \left(\frac{eB}{2{\pi}m_{e}c}\right)\gamma_e^{2}\frac{\Gamma}{1+z}$, where the magnetic field strength $B$ and the blast wave Lorentz factor $\Gamma$ can 
be estimated by the standard external shock model$^{21}$. Therefore, by calculating the Lorentz factor of the electrons producing the $R$ band emission, one can predict  
$P_{\rm cir}$ at the observing time as 
\[
P_{\rm cir} \sim10^{-4} \epsilon_{B,-2}^{1/4}E_{52}^{1/8}n^{1/8},
\]
where $\epsilon_{B} = 0.01\epsilon_{B,-2}$ is the fraction of the magnetic energy density to the internal energy 
density, $E = 10^{52}E_{52}$ [erg] is the total blast wave energy, and $n$ [cm$^{-3}$] is the circumburst particle number density. This value does not strongly depend on the model parameters, and is very low (in spite of the assumption that the magnetic field is ordered) compared to the observed value of $P_{\rm cir} = 0.61 \pm 0.13 \%$. 
In reality, the magnetic field directions are largely random, as implied by the observed $P_{\rm lin}$ light curve. However, the linear and circular polarization degrees are reduced to the same extent by the randomness of the field, so the ratio $P_{\rm cir}/P_{\rm lin} \sim 10^{-4}$ is applicable also for the random field case, which is clearly inconsistent with the observed value $P_{\rm cir}/P_{\rm lin} \sim 0.15$. 

In a situation where the pitch angle distribution is {\em not} isotropic, the circular polarization contributions of electrons are not cancelled out and $P_{\rm cir}$ can be 
higher. 
The observed  polarization ratio $P_{\rm cir}/P_{\rm lin}$, implies that $g(\theta)/\gamma_e \sim0.1$ and then  $g(\theta) \sim10^3$, which means a highly anisotropic
pitch angle distribution.

The detection of high circular polarization implies that the emitting plasma consists mainly of electrons and protons, rather than electrons and positrons, because the circular 
polarizations of the synchrotron emission of electrons and positrons perfectly cancel out$^{45}$. This implication is consistent with the emission model of the forward shock 
propagating in the circumburst medium.

\noindent [31] Barthelmy, S. D. {\it et al}. GRB 121024A: Swift-BAT refined analysis.  {\it GCN Circ}.  {\bf 13889}   
(2012)

\noindent [32] Tanvir, N. R. {\it et al}. GRB\,121024A: VLT/X-shooter redshift. {\it GCN Circ.} {\bf 13890} (2012) 

\noindent [33] Patat,  F. \& Romaniello, M. Error analysis for dual-beam optical linear polarimetry.  
 {\it PASP}  {\bf 118}, 146-161  (2006)

\noindent [34] Schlegel, D. J., Finkbeiner, D. P., Davis, M. Maps of dust infrared emission for use in
estimation of reddening and cosmic microwave background radiation foregrounds.  
 {\it Astrophys. J}.   {\bf 500}, 525-553 (1998)

\noindent [35] Serkowski, K., Matheson, D. S., Ford, V. L. Wavelength dependence of interstellar 
polarization and ratio of total to selective extinction.  {\it Astrophys. J}.   {\bf 196}, 261-290 (1975)

\noindent [36] Bagnulo, S., Szeifert, T., Wade, G. A., Landstreet, J. D., Mathys, G. Measuring magnetic 
fields of early-type stars with FORS1 at the VLT.  {\it Astron. Astrophys}.  {\bf 389}, 191-201 (2002)

\noindent [37] FORS Users Manual, Issue 91.1, Doc. No. VLT-MAN-ESO-13100-1543 (2012); available at \\
{\tt http://www.eso.org/sci/facilities/paranal/instruments/fors/doc.html}

\noindent [38] Wardle, J. F. C. \& Kronberg, P. P., The linear polarization of quasi-stellar radio sources at 
3.71 and 11.1 centimeters.  {\it Astrophys. J}.   {\bf 194},  249-255 (1974)

\noindent [39] Simmons, J. F. L. \&  Stewart, B. G. Point and interval estimation of the true unbiased 
degree of linear polarization in the presence of low signal-to-noise ratios.  {\it Astron. 
Astrophys}.  {\bf 142}, 100-106 (1985)

\noindent [40] Martin, P. G. Interstellar circular polarization.  {\it Mon. Not. R. Astron. Soc}.   {\bf 159}, 179-190 (1972)

\noindent [41] Klose, S. {\it et al}. Prospects for multiwavelength polarization observations of GRB afterglows 
and the case GRB 030329.   {\it Astron. Astrophys}.  {\bf 420}, 899-903 (2004)

\noindent [42] Greiner J. {\it et al}. GROND - a 7-Channel Imager.  {\it PASP}  {\bf 120}, 405 (2008)

\noindent [43] Hjorth, J. {\it et al}. The optically unbiased gamma-ray burst host (TOUGH) survey. I. Survey 
design and catalogs.  {\it Astrophys. J}.   {\bf 756}, 187-202 (2012)

\noindent [44] Pei, Y. C. Interstellar dust from the Milky Way to the Magellanic Clouds.  {\it Astrophys. J}.  
 {\bf 395}, 130-139 (1992)  

\noindent [45] Melrose, D. B.  {\it Non-thermal Processes in Diffuse Magnetised Plasmas}, Vol.  {\bf 1} 
(Gordon \& Breach, New York, 1980).

\newpage
\begin{figure}[h!]
\includegraphics[width=12cm]{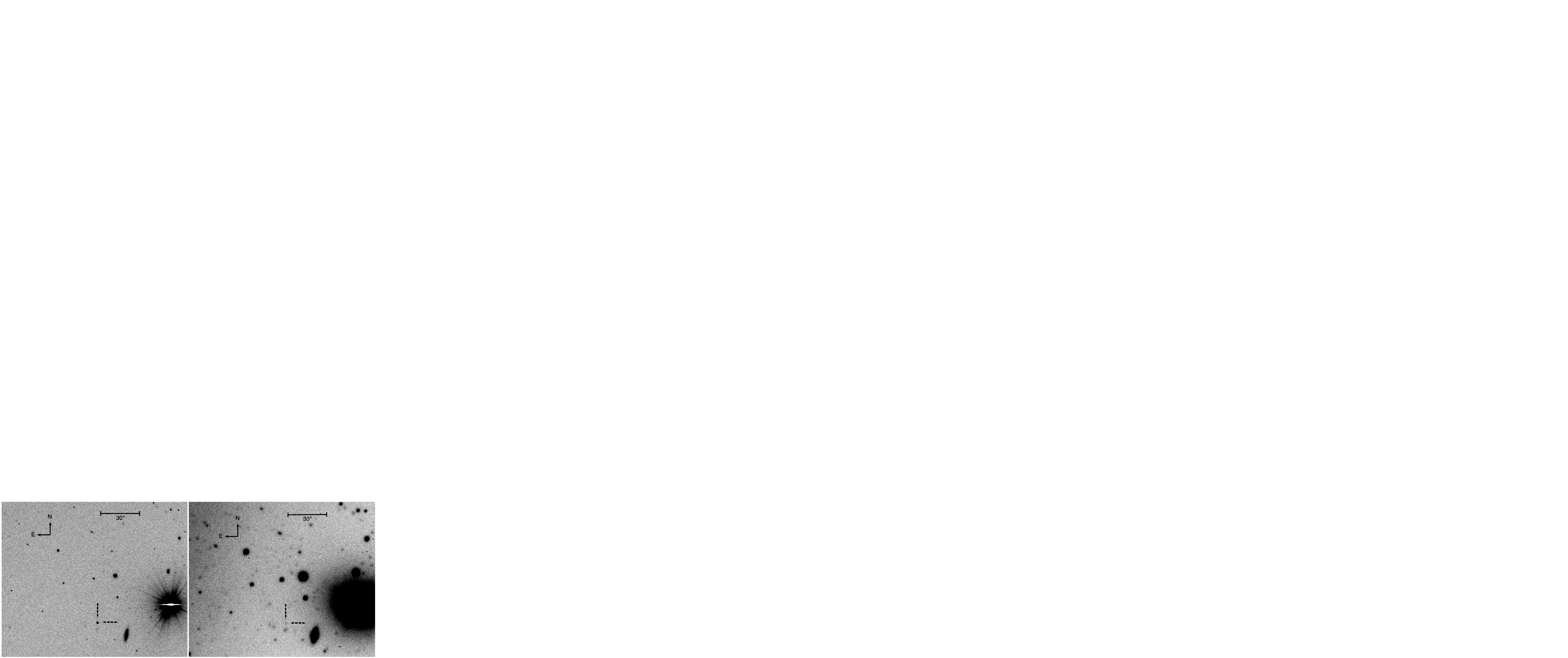}
\caption*{{\bf Extended Data Figure 1}. {\bf Host galaxy and afterglow image}. {\bf Left}, VLT FORS2 $R_{\rm special}$ band acquisition image, with the afterglow indicated by tick marks. {\bf Right}, Detection of the host galaxy in the late-time WHT ACAM $r$ band imaging.
}
\end{figure}

\begin{figure}[h!]
\includegraphics[width=12cm]{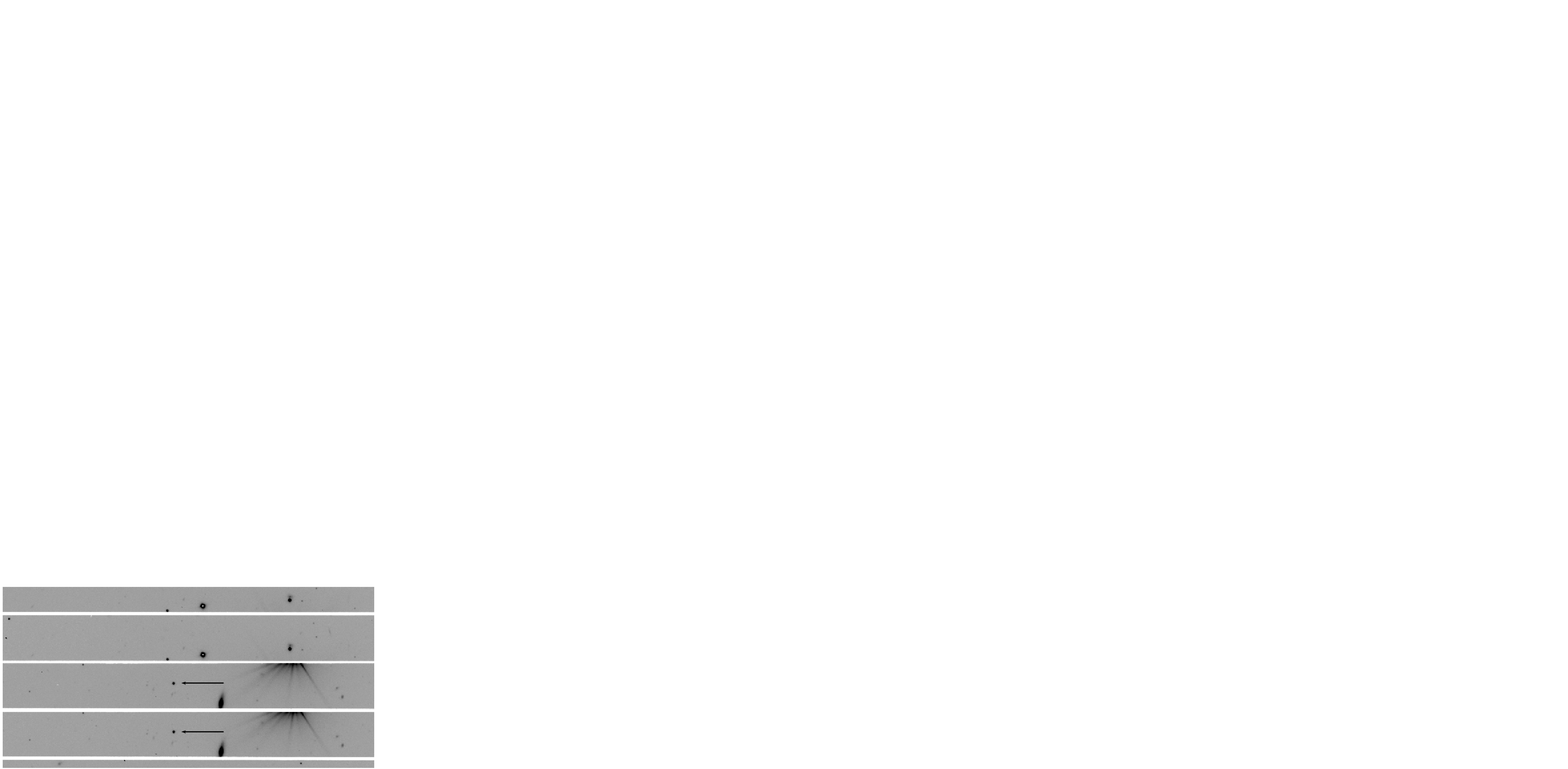}
\caption*{{\bf Extended Data Figure 2}. {\bf Polarimetry mask and afterglow brightness}.
A small section of a single FORS2 $R_{\rm special}$ band polarimetric exposure (this is the                
$-45^{\circ}$ angle chip 1 frame of the cir4 set), illustrating the shape of the aperture mask and brightness of the afterglow (indicated by an arrow).
}
\end{figure}

\begin{figure}[h!]
\includegraphics[width=12cm]{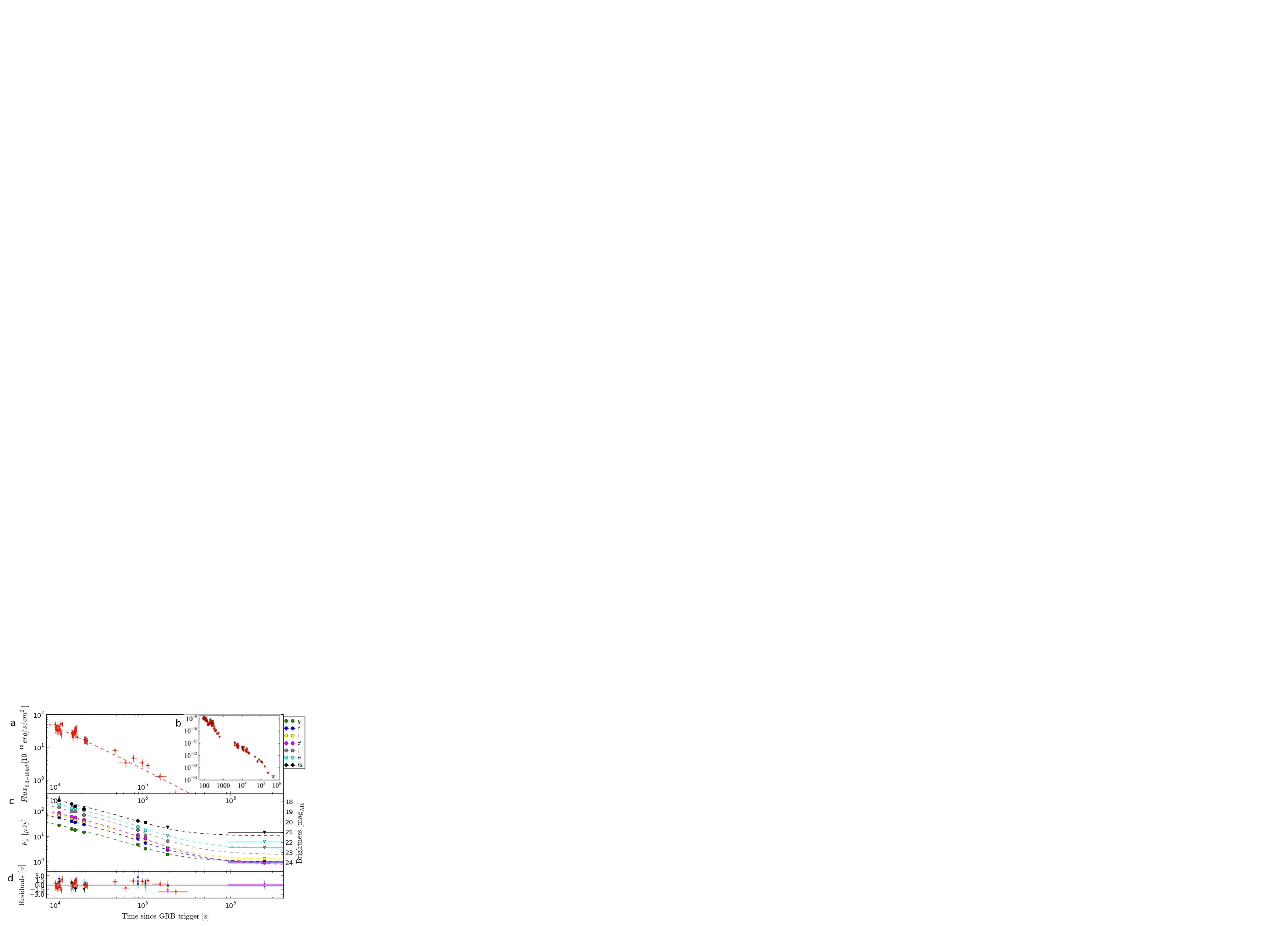}
\caption*{{\bf Extended Data Figure 3}. {\bf Optical and X-ray afterglow light curves}. 
{\bf a}, {\it Swift} XRT X-ray light curve in the time span covered by GROND observations; {\bf b}, The full XRT light curve. {\bf c}, Full GROND light curves in all
seven bands. Overplotted in panels {\bf a} and {\bf c} is the best fitting smoothly broken power law (Methods), with a host galaxy contribution to the optical data.  
Residuals to this fit are shown in {\bf d}.
}
\end{figure}

\begin{figure}[h!]
\includegraphics[width=10cm]{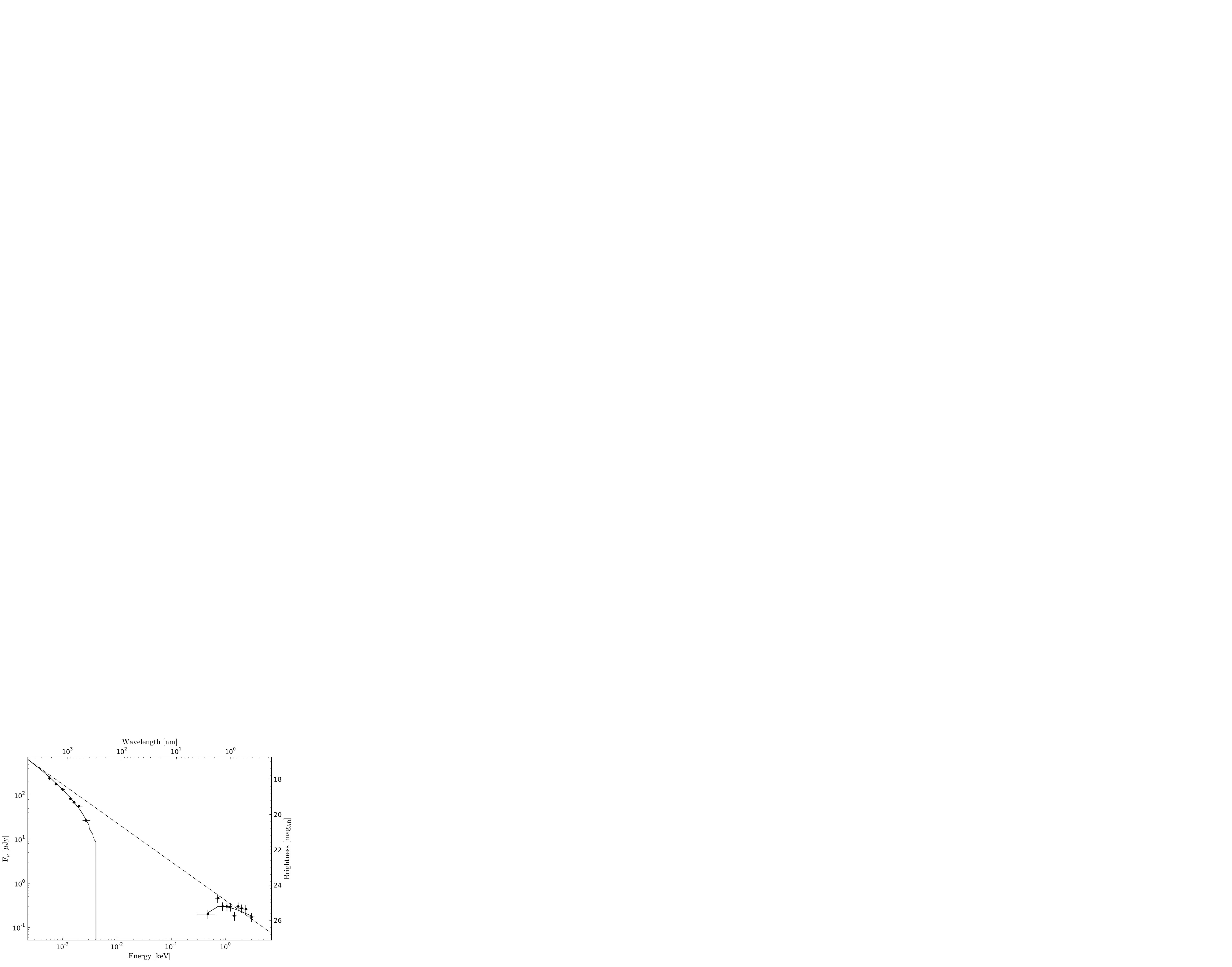}
\caption*{{\bf Extended Data Figure 4}. {\bf X-ray to optical spectral energy distribution of the afterglow of GRB\,121024A}. Shown is a spectral energy distribution 
using the seven GROND photometric bands and simultaneous Swift XRT X-ray data. The overplotted solid line is the best fitting absorbed power law; the dashed line shows the  
best fitting power law without the effects of reddening and X-ray absorption. 
}
\end{figure}

\begin{figure}[h!]
\includegraphics[width=12cm]{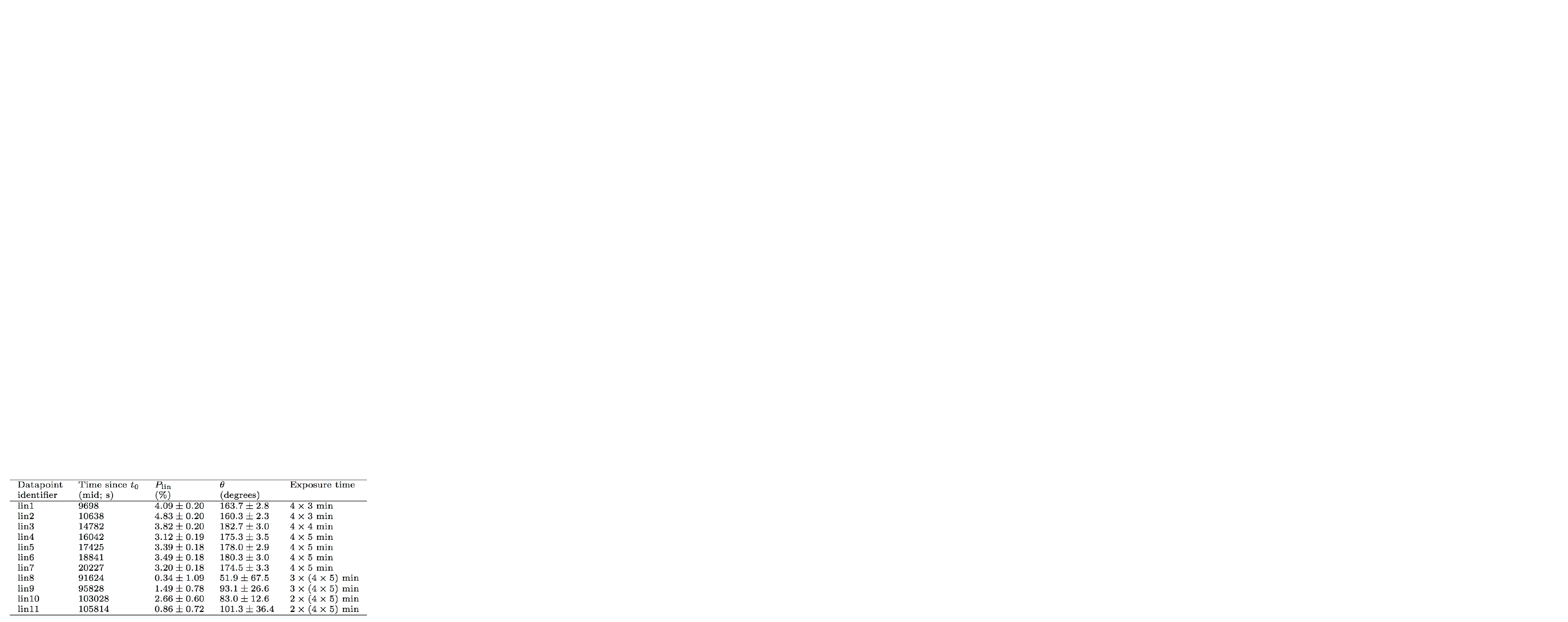}
\caption*{{\bf Extended Data Table 1}. {\bf Linear polarimetry results}. The polarization angle $\theta$ follows a standard coordinate convention: North=0$^{\circ}$, East=90$^{\circ}$. 
The values are corrected for Galactic dust induced polarization in Stokes parameter space, and polarization bias corrections are performed. Note that for the lin8 - lin11 
datapoints we combine multiple exposure sets together: the polarization is low and the source faint.
}
\end{figure}

\begin{figure}[h!]
\includegraphics[width=8cm]{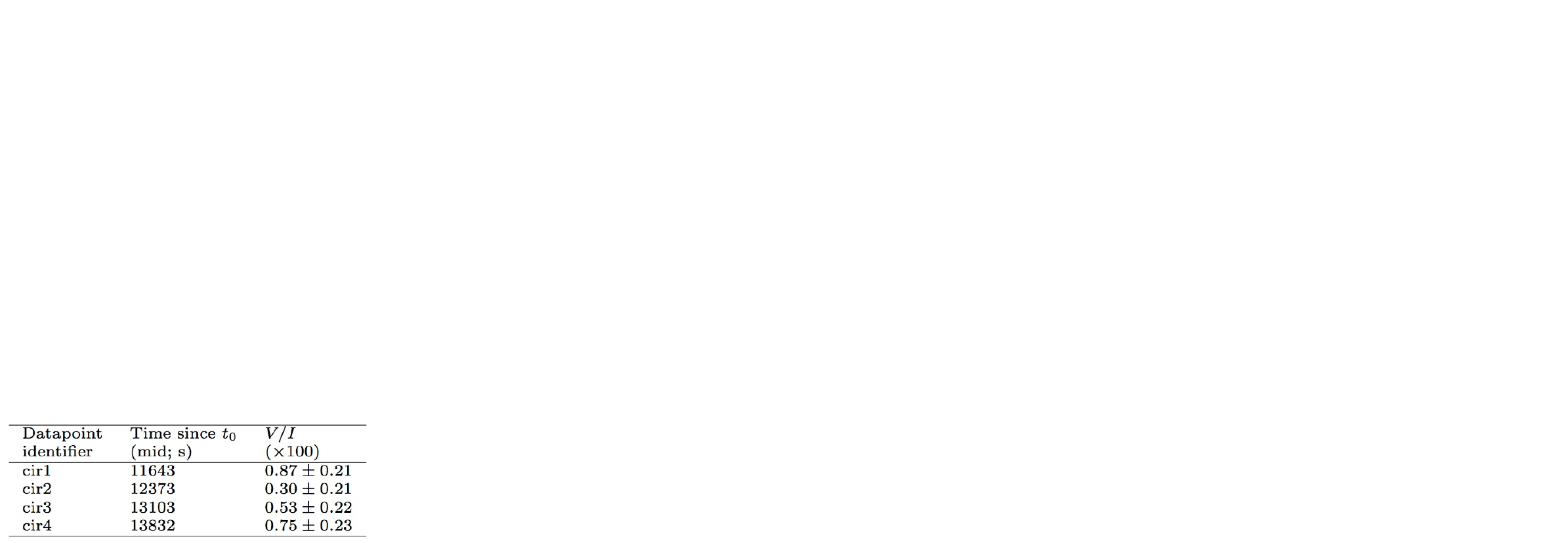}
\caption*{{\bf Extended Data Table 2}. {\bf Circular polarimetry results}. Each datapoint consists of two exposures ($-45^{\circ}$ and $+45^{\circ}$ angles) of 5 mins exposure time each. 
The sign of $V/I$ is positive for all four data points (the sign distinguishes clockwise and counterclockwise circular polarization direction), the circular polarization $P_{\rm cir}$ in 
percent is therefore equal to the values in the third column. Uncertainties are  $1\sigma$.
}
\end{figure}
\begin{figure}[h!]
\includegraphics[width=12cm]{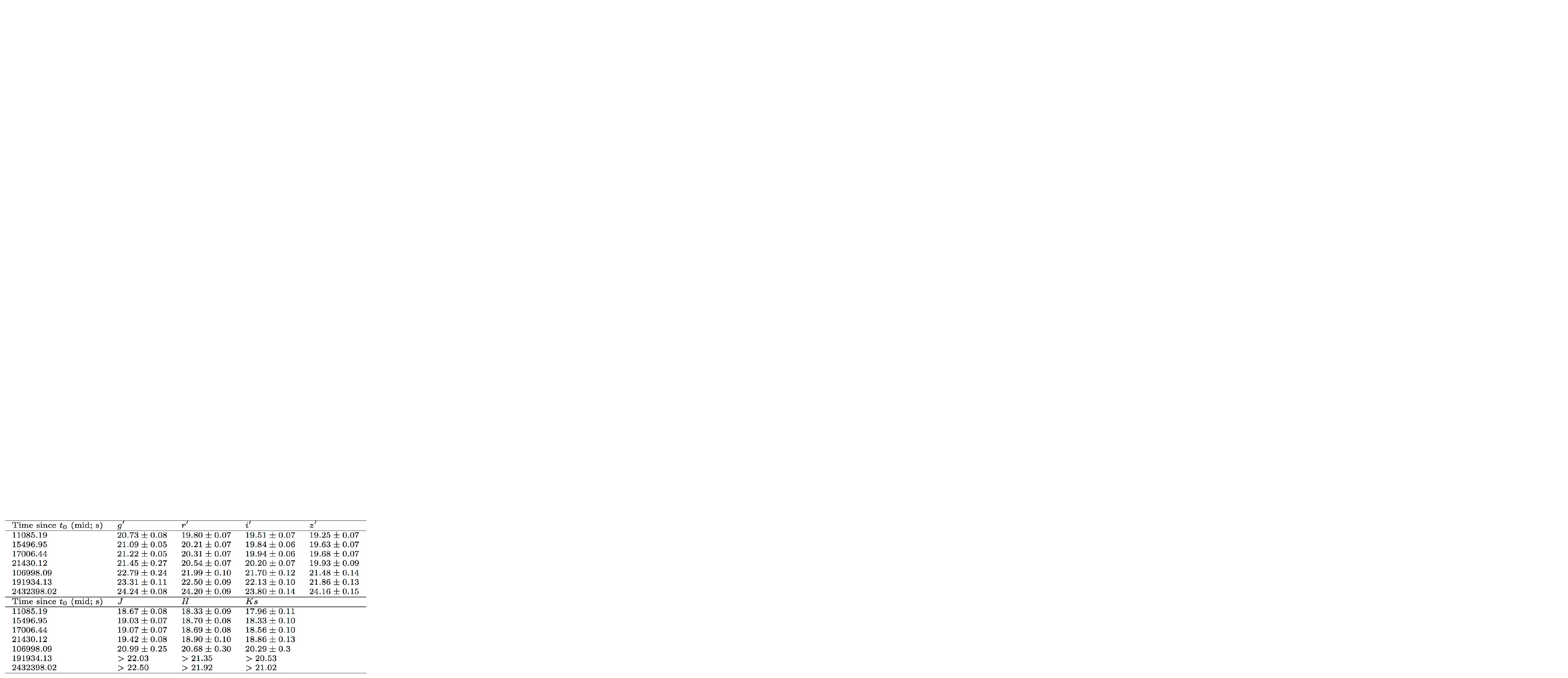}
\caption*{{\bf Extended Data Table 3}. {\bf GROND optical and near-infrared photometry of the afterglow}.  Magntitudes are as observed; uncertainties are  $1\sigma$, 
upper limits are $3\sigma$.
}
\end{figure}

\end{document}

%% file: supplementaryinfo.tex
\newenvironment{addendum}{%
    \setlength{\parindent}{0in}%
    \small%
    \begin{list}{Acknowledgments}{%
        \setlength{\leftmargin}{0in}%
        \setlength{\listparindent}{0in}%
        \setlength{\labelsep}{0em}%
        \setlength{\labelwidth}{0in}%
        \setlength{\itemsep}{12pt}%
        \let\makelabel\addendumlabel}
    }
    {\end{list}\normalsize}

\newcommand*{\addendumlabel}[1]{\textbf{#1}\hspace{1em}}

\begin{addendum}
\appendix
\item[Acknowledgments]
Based on observations made with ESO Telescopes at the Paranal Observatory under programme 090.D-0789. We thank all ING staff for their continuous support of ACAM ToO observations. KW thanks Jim Hinton for discussion. KW acknowledges support from STFC. KT acknowledges support from JSPS Research Fellowships for Young Scientists No. 231446. 
AJvdH, RAMJW and AR acknowledge support by the European Research Council via Advanced Investigator Grant no. 247295.
 RLCS is supported by a Royal Society Fellowship. YZF is supported by the 973 Programme of China, under grant 2013CB837000.
 DMR acknowledges support from a Marie Curie Intra European Fellowship within the 7th European Community Framework Programme under contract no. IEF 274805. This work was supported by Australian Research Council grant DP120102393. The William Herschel telescope and its override programme are operated on the island of La Palma by the Isaac Newton Group in the Spanish Observatorio del Roque de los Muchachos of the Instituto de Astrof\'{i}sica de Canarias. This work made use of data supplied by the UK {\it Swift} Science Data Centre at the University of Leicester, funded by the UK Space Agency.

\item[Author Contributions]
 KW and SC jointly led the VLT observing time proposals and defined the observing strategy. KW acquired, reduced and analysed the VLT data and took primary responsibility for writing the text of the paper; SC performed an independent data analysis. KT, AvdH and MM provided theoretical interpretation of the observations. KV and JG analysed the GROND data. OH led the WHT observing time proposal. All authors contributed to refining the text of the paper, or assisted in obtaining parts of the presented dataset.
 \item[Competing Interests] The authors declare that they have no
competing financial interests.
 \item[Correspondence] Correspondence and requests for materials
should be addressed to KW (kw113@le.ac.uk)




\end{addendum}